# Coherence properties of light propagated through a scattering medium

C.K. Aruldoss*, N. Dragomir, K.A. Nugent and A. Roberts
School of Physics, University of Melbourne, Victoria-3010, AUSTRALIA

**ABSTRACT**

Partially-coherent, quasi-monochromatic optical fields are fully described by their Mutual Optical Intensity (MOI) or the phase-space equivalent, the Generalised Radiance (GR). This paper reports on the application of a propagation-based phase-space tomographic technique for determining both the MOI and the GR of wavefields. This method is applied to the reconstruction of the MOI and the GR of an optical wavefield propagated through a suspension of ~10 μm diameter polystyrene spheres.

## 1. INTRODUCTION

The spatial coherence properties of wavefields are of fundamental importance since they play a key role in many optical imaging techniques.[1] For example, it is well-known that partially coherent wavefields have the potential to carry more information content than fully coherent wavefields and that the use of partially coherent illumination in imaging can yield higher spatial resolution.[2] Hence, the characterisation of the coherence properties of a wavefield embodied in the MOI is of critical importance for applications including imaging and lithography. In addition, the spatial coherence properties of a lightfield evolve as it propagates through a scattering medium such as biological tissue.[3] Knowledge of these changes in an optical field is critical to a complete understanding of imaging and the development of new imaging methods. In particular, once the MOI of an optical field has been determined, it can be used to predict the field in any arbitrary plane. In particular, provided the optical properties of the medium through which has propagated are well-understood, the field can be 'back-propagated' to a plane of interest. This property of the MOI, therefore, can form the basis of new forms of imaging. The research presented here is aimed at developing new non-interferometric techniques for fully characterising partially coherent optical fields and their application to problems in imaging.

Traditionally, measurements of the spatial coherence of a source have been obtained using double-slit experiments[4] or uniformly redundant arrays.[5] These techniques, however, provide only limited information, for example for certain slit separations at isolated points in the source. Various alternative interferometric techniques for determining the MOI of a wavefield have also been proposed and demonstrated[6], but these are generally complex and require more precise alignment than non-interferometric methods. The technique used here is based on a non-interferometric method originally proposed by Nugent[7] and extended by Raymer et al.[8] to two dimensional fields. Subsequently, McAlister et al.[9] effectively reconstructed the wave field of partially coherent sources using a fractional Fourier transform based algorithm. Anhut et al.[10] also reconstructed the space coherence function of the forward scattering field for large diameter (90μm) spheres employing phase-space tomography using the Radon transformation. The numerical aperture of their system, however, was extremely small and limited extension of this method to media containing smaller scatterering particles. Here we report on the development of a non-interferometric phase-space technique for recovering both the MOI and the GR of an optical wavefield and present preliminary results demonstrating its application to the characterisation of the optical field after propagation through a medium containing polystyrene spheres of approximately 10μm diameter.

## 2. PHASE-SPACE TOMOGRAPHY

**1.1 Cross-correlation functions**

Quasi-monochromatic spatially, partially coherent optical fields can be described in terms of the correlation between the field at two points $\vec{r}_1$ and $\vec{r}_2$ using the MOI, $G$:

$$G(\vec{r}_1, \vec{r}_2) = \left\langle E(\vec{r}_1) E^*(\vec{r}_2) \right\rangle \tag{1}$$

where $E(\vec{r})$ represents the complex field amplitude at position $\vec{r}$, the asterisk represents complex conjugation and the brackets $<>$ represent an average over all realisations of the field. Alternatively, the phase-space representation of the wavefield is written in terms of the GR:

$$B(\vec{r},\vec{u}) = \frac{1}{(2\pi)^2} \int J(\vec{r},\vec{\Delta}) e^{-i\vec{\Delta}\bullet\vec{u}} d\vec{\Delta} \qquad (2)$$

where $\vec{r} = (\vec{r}_1 + \vec{r}_2)/2$, $\vec{\Delta} = \vec{r}_1 - \vec{r}_2$, $\vec{u}$ is the momentum coordinate conjugate to $\vec{\Delta}$ and $J(\vec{r},\vec{\Delta}) = G(\vec{r}+\vec{\Delta}/2, \vec{r}-\vec{\Delta}/2)$. The reconstruction of the MOI or GR gives us a way to fully characterize the wavefield. For example, the intensity, $I(\vec{r})$ at a point $\vec{r}$ is given by $J(\vec{r},0) = G(\vec{r},\vec{r})$ or equivalently as a projection over the momentum coordinate of the GR:

$$I(\vec{r}) = \int B(\vec{r},\vec{u}) d\vec{u}. \qquad (3)$$

The cross-correlation function, the MOI fully describes the properties of the field, while the GR function contains the same information, but represents the field in phase-space.

**1.2 Theory**
The process of reconstructing the field by means of intensity measurements at different planes corresponding to different angles of observation is termed phase-space tomography. It can be shown[7] that in the paraxial approximation as a field propagates a distance $z$ its GR is 'sheared' in phase space:

$$B(\vec{r},\vec{u},z) = B\left(\vec{r} - \frac{z\vec{u}}{k}, \vec{u}, 0\right), \qquad (4)$$

where $k = \frac{2\pi}{\lambda}$ is the wavenumber of the field and $\lambda$ is its free-space wavelength. Since the intensity at a plane $z$ represents a projection through momentum (equation (3)), measurements of intensity as a field propagates provide a series of projections through the GR. This is the fundamental basis of phase-space tomography since these projections can be used to reconstruct the GR and, hence, the MOI.

The technique to be used here, Phase Space Tomography, involves measuring the intensity $I(\vec{r},z)$ of the field after passing through a lens of focal length $f$ as a function of distance $z$ from the lens to reconstruct the wavefield at some distance $z_1$ behind the lens (Figure 1). In this case, it can be shown that the relationship between the Fourier transform of the intensity distribution, $\tilde{I}(\vec{\rho},z)$, and the Fourier transform of the GR at the plane of interest, $\tilde{B}(\vec{u},\vec{v})$, are related by the equation:

$$\tilde{I}(\vec{\rho},z) = \tilde{B}\left(\vec{\rho}\left(1-\frac{z}{f}\right), \frac{\vec{\rho}}{k}\left(z_1 - z - \frac{z_1 z}{f}\right)\right) \qquad (5)$$

where $\vec{\rho}$ is the coordinate conjugate to position $\vec{r}$. According to Equation (5), in the case of a one-dimensional field we can deduce the existence of a one-to-one correspondence between the Fourier transforms of the intensity distribution for a wave propagated to a distance $z$ from the lens and a line through the Fourier transform of the GR in the plane of interest.

For one-dimensional fields, i.e. fields that depend on only one transverse coordinate, measurements of the intensity as a function of $z$ are sufficient to completely determine the Fourier transform of the GR, and hence the GR and the MOI, in the plane $z = 0$. Note that for one-dimensional fields, the MOI and the GR are two-dimensional functions. Using this procedure, it is also possible to completely reconstruct the MOI and the GR of separable fields, i.e. fields that can be written in the form:

$$J(\vec{r}, \vec{\Delta}) = X(x, \Delta_x) Y(y, \Delta_y) \qquad . \qquad (6)$$

Examples of fields of this kind include Gaussian beams. In the general case of two-dimensional fields, three-dimensional intensity measurements are sufficient to reconstruct projections through the four-dimensional MOI and, hence, still provide comprehensive and valuable information about the coherence and other properties of the field.

Note that since the MOI can theoretically be reconstructed at all points, the mutual optical intensity and the spatial coherence function can be obtained at all points for all separations $\vec{\Delta}$. This would correspond to obtaining information from a series of double-slit experiments where the coherence function was measured when the slits were placed at all points within the field of interest and the slits were adjusted for all separations. Hence, phase-space tomography has the capacity to obtain much more comprehensive information about a wavefield than traditional non-interferometric techniques.

## 3. EXPERIMENT

The experimental set-up used to obtain the MOI and the GR of the wavefield is shown in Figure 1. The light radiated from a single-mode fibre-coupled Superluminescent Diode (SLD) (Superlum Ltd. Model 505) with a central wavelength of 681 nm and a linewidth of 12 nm was collimated using a microscope objective of magnification 10× and transmitted through a lens of focal length 15 cm.

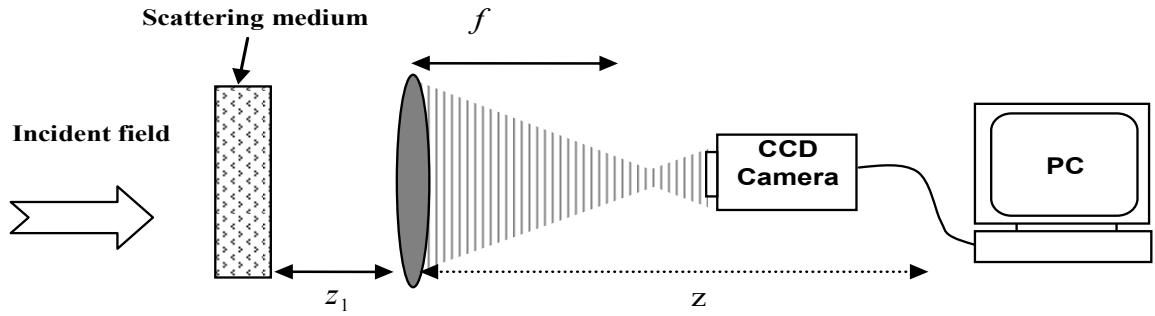

Figure 1: Schematic diagram showing the principle of non-interferometric phase-space tomography for wavefield recovery.

The purpose of the experiment is to measure the intensity distributions and hence the Fourier transform of the GR for the entire range of angles. The value of the angle of observation can be changed (see Equation (5)) by varying either the focal length ($f$) of the lens or the distance between the lens and CCD camera ($z$). In this experiment the second possibility is utilized with the focal length fixed at 15 cm. The propagated light field was recorded using a 1340 × 1300 pixel 16-bit Versarray CCD camera (Roper Scientific Inc.) with 20 μm pixel size controlled by a computer. Intensity images were recorded at regular intervals as the camera was moved along a fixed rail Fig. (1). Each image was obtained with an exposure of 10 ms while the camera temperature was maintained at -20° C. Images were collected over values of $z$ ranging from 4.0 cm to 33.8 cm. The minimum value of $z$ was dictated by the geometry of the CCD camera.

The reconstruction of the GR and the MOI of the light after propagation through a scattering medium was performed by obtaining intensity measurements using the same technique. In this case polystyrene microspheres (Polyscience Inc.) of density 1.06 g/cm$^3$ in water were used as the scattering medium. A cuvette of path length 1 mm was filled with a dilute water (80%)/glycerol (20%) solution containing microspheres of diameter 10 μm ± 1 μm with a number density of 42×10$^6$ /cm$^3$. The cuvette was introduced into the collimated beam between the source and the lens a distance of 21.5 cm from the lens as shown in Figure 1 and the three-dimensional intensity distribution was obtained in the same manner as discussed above. To obtain an ensemble average, the sample was stirred between, but not during, exposures and three images were averaged at each position.

## 4. RESULTS AND DISCUSSION

The GR and the MOI of the incident unscattered and the scattered wavefields described above were reconstructed using the theory described in the previous sections. Figure 2 shows the modulus of the reconstructed MOI for the unscattered beam (a) and the beam after it has passed through the scattering medium (b). The MOI is also represented as a surface plot in the form $J(r,\Delta)$ in Figure 3.

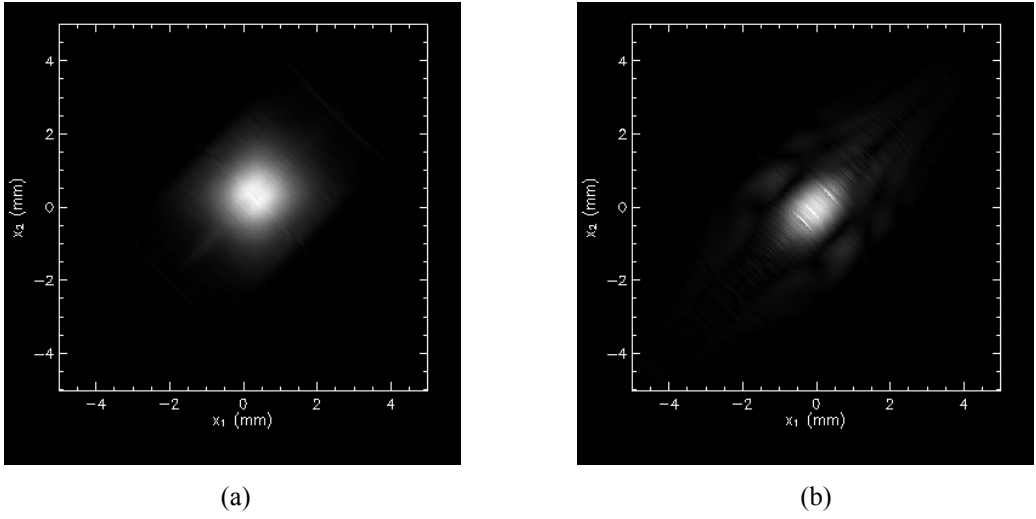

(a)                      (b)

Figure 2: The reconstructed MOI, $G(x_1, x_2)$ for a collimated beam (a) and the same beam after passing through the scattering medium (b).

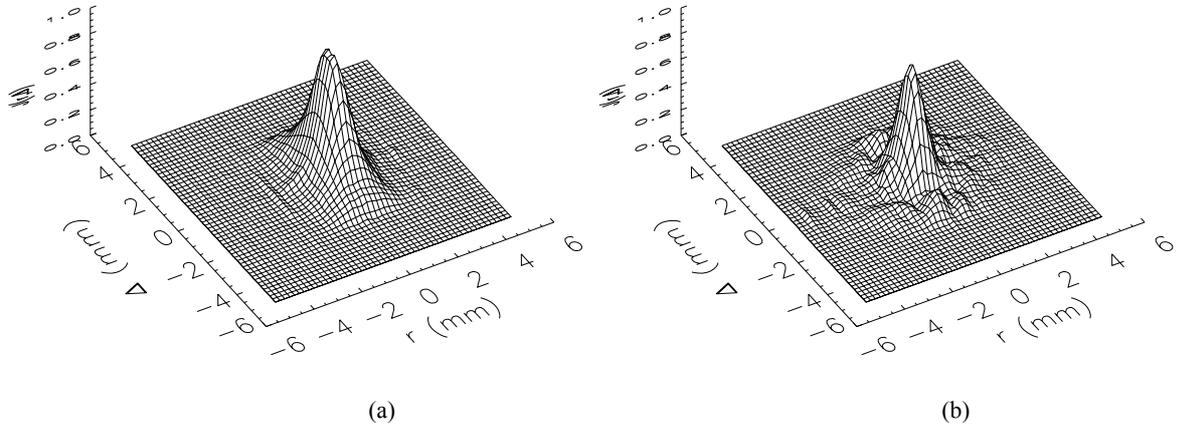

(a)                      (b)

Figure 3: The normalised reconstructed MOI, $J(r,\Delta)$ for a collimated beam (a) and the same beam after passing through the scattering medium (b).

When partially coherent light propagates through a strongly forward-scattering medium, its GR, $B_{tot}$, can be regarded as being the sum of three terms:[11]

$$B_{tot} = B_{ballistic} + B_p + B_b. \qquad (8)$$

The first term describes the unscattered, ballistic component of the optical field, the second the contribution from the near-forward Mie scattered light and the third term the broad angle contribution. The experiments performed here have the potential to determine the contribution to the GR from the first two terms. The second term adds a broader pedestal background to the unscattered narrow angle term in the GR. When considering the MOI, on the other hand, the near-forward scattering is expected to introduce a more sharply peaked lower coherence contribution around $\vec{\Delta} = 0$ in addition to the broader ballistic component.

From Figures 2 and 3, it can clearly be seen that the MOI has been changed by the presence of scattering and, as predicted, there is a narrower scattered component in the reconstructed MOI superposed on a weak ballistic component. Note that the overall width of the ballistic component is the same in the presence of scattering as in its absence. The ability to discriminate between the ballistic and scattered contributions of an optical wavefield could form the basis of new coherence gating techniques in imaging.

Finally, the phase space representation, the GR, is shown in Figure 4. In Figure 4(a) the GR of the unscattered beam is shown. It can be seen that in phase-space the angular (vertical) spread is relatively narrow. After the beam has passed through the scattering medium, however, there has been a broadening in the GR as a consequence of broader angle near-forward scattering.

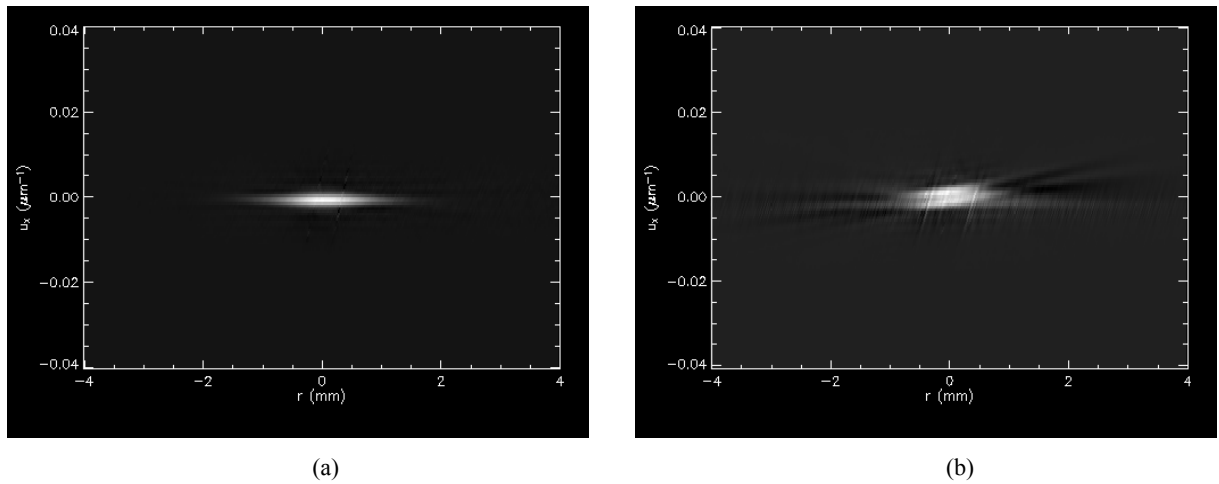

(a)  (b)

Fig: 4. The reconstructed GR, $B(r,u)$ for a collimated beam (a) and the same beam after passing through the scattering medium (b).

One of the challenges of any tomographic procedure is the development of procedures to process incomplete information. In this case, images were obtained over only a finite range of $z$ values which limited the range of accessible projections, although far-field extrapolation was used where appropriate. The mapping of the Fourier transform of the GR was also constrained by the resolution and size of the CCD chip in the camera. These limitations led to the appearance of reconstruction artefacts in the computed MOI and GR and the development of improved algorithms is the subject on ongoing work. Another issue of importance is the fact that the scattering medium should be constantly agitated during image exposures to ensure the ensemble averaging implicit in the definition of the MOI defined in equation (1). The results shown here were obtained with no stirring during exposures and averaging only three images and, hence, the structure evident in the broader ballistic component is believed to be a consequence of not performing sufficient averaging. Obtaining more consistent ensemble averaged information is another goal of this research.

## CONCLUSION

A non-interferometric phase-space tomographic method for the characterisation of the spatial coherence properties of optical fields has been presented and its application to the characterisation of wavefields that have passed through a scattering medium demonstrated. This non-interferometric method is straightforward to implement and, unlike other methods involving slits or uniformly redundant arrays, provides complete information about a quasi-monochromatic lightfield including a measure of the spatial coherence properties of the field at each point in the wavefield. Although research in this topic is ongoing, the preliminary results presented here demonstrate the utility of the technique. Further work will be aimed at developing enhancements to the experimental arrangements, improving the reconstruction algorithms, performing theoretical calculations and simulations and undertaking comprehensive studies of the influence of scattering on the coherence of light.

## ACKNOWLEDGEMENTS

The authors would like to acknowledge the financial support of the University of Melbourne Research Grant Scheme and the Australian Research Council. KAN is the recipient of an Australian Research Council Federation Fellowship and CKA the recipient of an Australian International Postgraduate Research Scholarship and a University of Melbourne International Research Scholarship. The authors would also like to thank Chanh Tran, Robert Scholten and Adrian Mancuso for helpful discussions.

*\*celine@physics.unimelb.edu.au; phone +61 (0)3 8344 0144*